\documentclass[10pt]{article}
\usepackage{amssymb}
\usepackage{graphics}
\usepackage{epsfig}
\usepackage{a4wide}
\usepackage{slashed,cite}
\usepackage{color}

\begin{document}
\title{  $T$-odd quark pair production and decay at $\gamma \gamma$ collider in the littlest Higgs model with $T$ parity in next-to-leading order QCD}
\author{  Liu Wen,$^1$ Ma Wen-Gan,$^1$ Guo Lei,$^2$ Chen Liang-Wen,$^1$ Chen Chang,$^1$ and Zhang Ren-You$^1$ \\
$^1${\small  Department of Modern Physics, University of Science and Technology of China (USTC), }  \\
{\small   Hefei, Anhui 230026, P.R.China} \\
$^2${\small  Department of Physics, Chongqing University, Chongqing 401331, P.R. China}
}

\date{}
\maketitle \vskip 15mm
\begin{abstract}
We calculate the complete next-to-leading order (NLO) QCD corrections to the $T$-odd mirror quark pair ($q_-\bar q_-=u_-\bar u_-$, $c_-\bar c_-$, $d_-\bar d_-$, $s_-\bar s_-$) production in the littlest Higgs model with $T$-parity (LHT) at a high energy $\gamma \gamma$ collider. We present the dependence of the leading order (LO) and NLO QCD corrected cross sections on the colliding energy $\sqrt{s}$. Our calculation includes the subsequent full weak decays of the final $T$-odd mirror quarks by adopting the narrow width approximation and the exclusive 2-jet event selection criterion. We provide the LO and QCD NLO kinematic distributions of final particles. We find that the NLO QCD correction is phase space dependent and modifies the LO cross section evidently. The $K$-factor increases noticeably when $\sqrt{s}$ approaches the threshold of the on-shell $q_-$-pair production. We conclude that it is possible to separate the signature of the $T$-odd quark pair production from possible Standard Model background by taking proper kinematic cut.
\end{abstract}

\vskip 15mm {\bf PACS: 12.38.Bx, 13.85.Lg, 13.85.Hd  }

\vskip 5mm

{\bf Keywords: Littlest Higgs model with $T$ parity, $T$-odd quark pair production, photon-photon collider }

\vfill \eject \baselineskip=0.32in

\renewcommand{\theequation}{\arabic{section}.\arabic{equation}}
\renewcommand{\thesection}{\Roman{section}.}
\newcommand{\nb}{\nonumber}

\newcommand{\Dir}{\kern -6.4pt\Big{/}}
\newcommand{\Dirin}{\kern -10.4pt\Big{/}\kern 4.4pt}
\newcommand{\DDir}{\kern -7.6pt\Big{/}}
\newcommand{\DGir}{\kern -6.0pt\Big{/}}

\makeatletter      
\@addtoreset{equation}{section}
\makeatother       

\par
\section{Introduction }
\par
Recently the discovery of a new particle that resembles the Standard Model (SM) Higgs boson was announced simultaneously by the CMS and ATLAS collaborations at CERN Large Hadron Collider\cite{Aad:2012tfa,Chatrchyan:2012ufa}. This amazing discovery is an important milestone of physics and has given an impetus to the study of the electroweak symmetry breaking (EWSB) mechanism. Subsequent analyses are still in process to determine the traits of the new boson.

\par
Although the SM has made immense success in explaining fundamental particle interactions, there are still crucial questions, such as the famous hierarchy problem and the search for new physics. In order to solve the hierarchy problem in the SM, several models have been proposed. Among them the little Higgs mechanism \cite{LTM-1,LTM-2,LTM-3} is one of the elegant candidates. The littlest Higgs model (LHM) is the most economical realization of the little Higgs mechanism and a phenominologically viable model. In the LHM, a scalar triplet $\Phi$, a new vector-like quark $T$ and a set of new heavy gauge bosons ($W_H^{}$, $A_H$, $Z_H$) are introduced. However, the mixture of the SM gauge bosons and the heavy new gauge bosons leads to stringent constraints on the LHM. Motivated by these constraints, a new discrete symmetry, denoted as ``$T$-parity'', is introduced\cite{T-parity-1,T-parity-2,Low:2004xc,Hubisz:2004ft,Hubisz:2005tx}. In the littlest Higgs model with $T$-parity (LHT), all the SM particles are $T$-even while all new particles are $T$-odd except $T_+$. As a result of the $T$-parity conservation, the mixture between the SM gauge bosons and the heavy new gauge bosons is prohibited and no triplet vaccum expectation value (VEV) is generated. Consequetly, the strong electroweak constraints on the model are significantly relaxed. In the LHT, a neutral, colorless and weakly interacting stable particle $A_H$ is predicted and can be a good candidate for dark matter \cite{Low:2004xc,Hubisz:2004ft,Hubisz:2005tx,Barbieri:2000gf,Cheng:2003ju}, which has gained a lot of attention.

\par
The CERN Large Hadron Collider can directly produce very massive new particles and will extend the possibilities of finding new physics effects, but it cannot easily provide precision measurements due to the typical characteristic of hadron machine. Whereas, a TeV scale linear collider with extremely high luminosity and clean experimental environment, can provide complementary information on these properties with precision measurements that would complete the results from hadron collider experiment. A most popularly proposed linear collider with energies at TeV scale and extremely high luminosity is the Compact Linear Collider (CLIC) \cite{Braun,Abramowicz}. In addition to $e^+e^-$ collision, linear collider provides a suitable platform running in $\gamma\gamma$ and $\gamma e$ collision modes at energies and luminosities comparable with those in $e^+e^-$ collision mode through the laser backscattering procedure \cite{Ginzburg-1,Ginzburg-2}.

\par
The LHT phenomenology of the $T$-odd quarks has been extensively studied. The effects of the $T$-odd fermions in high energy scattering processes were explored in Ref.\cite{LHT-ph1}. Recently, the $T$-odd quark signals at the LHC and ILC were studied at the QCD NLO in Refs.\cite{LHT-ph21,LHT-ph22,LHT-ph23,LHT-ILC}, and the QCD NLO correction to the $T$-odd gauge boson associated with a $T$-odd quark production at the LHC was reported in Ref.\cite{LHT-ph41,LHT-ph42}.

\par
We investigate the $q_-$-pair production up to the QCD NLO in the LHT at the future high energy $\gamma \gamma$ collider, including subsequent decays of the final $T$-odd quarks in this paper. The content is organized  as follows: In Sec. II, we briefly review the related LHT theory, and provide the corresponding mass spectrum. In Sec. III, the calculation strategy is presented. In Sec. IV, numerical analysis and discussion are provided. The last  section is devoted to the summary.

\vskip 3mm
\section{Related LHT theory and mass spectrum}
\par
The LHT is a nonlinear $\sigma$-model based on the $SU(5)/SO(5)$ global symmetry breaking at some high energy scale $f$, leading to 14 massless Nambu-Goldstone bosons described by the ``$\Pi$'' matrix as
\begin{eqnarray}
\label{pion matrix}
\Pi= \left(
\begin{array}{ccccc}
-\frac{\omega^0}{2} - \frac{\eta}{\sqrt{20}} &
-\frac{\omega^+}{\sqrt{2}} &
-i\frac{\pi^+}{\sqrt{2}} &
-i\phi^{++} &
-i\frac{\phi^+}{\sqrt{2}} \\
-\frac{\omega^-}{\sqrt{2}} &
\frac{\omega^0}{2} - \frac{\eta}{\sqrt{20}} &
\frac{v+h+i\pi^0}{2} &
-i\frac{\phi^+}{\sqrt{2}} &
\frac{-i\phi^0+\phi^P}{\sqrt{2}} \\
i\frac{\pi^-}{\sqrt{2}} &
\frac{v+h-i\pi^0}{2} &
\sqrt{4/5}\eta &
-i\frac{\pi^+}{\sqrt{2}} &
\frac{v+h+i\pi^0}{2} \\
i\phi^{--} &
i\frac{\phi^-}{\sqrt{2}} &
i\frac{\pi^-}{\sqrt{2}} &
-\frac{\omega^0}{2}-\frac{\eta}{\sqrt{20}} &
-\frac{\omega^-}{\sqrt{2}} \\
i\frac{\phi^-}{\sqrt{2}} &
\frac{i\phi^0+\phi^P}{\sqrt{2}} &
\frac{v+h-i\pi^0}{2} &
-\frac{\omega^+}{\sqrt{2}} &
\frac{\omega^0}{2} - \frac{\eta}{\sqrt{20}}
\end{array}\right).
\end{eqnarray}
Among the 14 Nambu-Goldstone bosons, $\eta$, $\omega^0$ and $\omega^{\pm}$ are the Goldstone bosons associated with the spontaneous gauge symmetry breaking
\begin{eqnarray}
\left[SU(2) \otimes U(1) \right]_1 \otimes\left[SU(2) \otimes U(1) \right]_2 \to SU(2)_L \otimes U(1)_Y,
\end{eqnarray}
and are eaten by the heavy gauge bosons $A_H$, $Z_H$ and $W^{\pm}_H$, respectively. The other 10 Nambu-Goldstone bosons constitute two $SU(2)_L$ multiplets: (1) $T$-even SM Higgs doublet $H \sim \left(\pi^{+}, h+v, \pi^0\right)$, where $h$ is the SM Higgs boson, $v \simeq 246~ {\rm GeV}$ the Higgs VEV, and $\pi^{0, \pm}$ are the Goldstone bosons eaten by the SM gauge bosons, and (2) $T$-odd scalar triplet $\Phi \sim \left(\phi^{++}, \phi^{+}, \phi^0, \phi^P\right)$.

\par
The $\left[SU(2) \otimes U(1) \right]_1 \otimes\left[SU(2) \otimes U(1) \right]_2$ gauge fields $B_i$ and $W_i^a$ $(i = 1, 2,~ a = 1, 2, 3)$ transform under $T$ parity as
\begin{eqnarray}
B_1 \longleftrightarrow B_2,~~~~~~~~~ W_1^a \longleftrightarrow W_2^a.
\end{eqnarray}
The heavy gauge bosons $A_H$, $Z_H$ and $W^{\pm}_H$ are the $T$-odd eigenstates of the $\left[SU(2) \otimes U(1) \right]_1 \otimes\left[SU(2) \otimes U(1) \right]_2$ gauge fields, and therefore can be expressed as
\begin{eqnarray}
&& \left(
\begin{array}{c}
A_H \\
Z_H
\end{array}
\right)
=
\left(
\begin{array}{cr}
\cos\theta_H & -\sin\theta_H \\
\sin\theta_H & \cos\theta_H
\end{array}
\right)
\left(
\begin{array}{cccc}
\frac{1}{\sqrt{2}} & -\frac{1}{\sqrt{2}} & 0 & 0 \\
0 & 0 & \frac{1}{\sqrt{2}} & -\frac{1}{\sqrt{2}}
\end{array}
\right)
\left(
\begin{array}{c}
B_1 \\
B_2 \\
W_1^3 \\
W_2^3
\end{array}
\right), \nonumber \\
&& W_H^{\pm}
=
\frac{\left(W_1^1 - W_2^1\right) \mp i \left(W_1^2 - W_2^2\right)}{2}.
\end{eqnarray}
At ${\cal O}(v^2/f^2)$, the masses of the $T$-odd heavy gauge bosons are given by
\begin{eqnarray}
\label{VH mass}
m_{A_H} = \frac{1}{\sqrt{5}} g^{\prime} f \left( 1 - \frac{5}{8}\frac{v^2}{f^2} \right),~~~~~~~
m_{Z_H} = m_{W_H} = g f \left( 1 - \frac{1}{8}\frac{v^2}{f^2} \right),
\end{eqnarray}
and the mixing angle $\theta_H$ has the form as
\begin{eqnarray}
\sin\theta_H = \frac{5 g g^{\prime}}{4 (5 g^2 - g^{\prime 2})} \frac{v^2}{f^2},
\end{eqnarray}
where $g$ and $g^{\prime}$ are the $SU(2)_L$ and $U(1)_Y$ gauge couplings, respectively.

\par
A consistent implementation of $T$ parity in the quark sector requires the introduction of the $T$-odd mirror quarks for the SM quarks. For each quark flavor, we introduce the following two incomplete left-handed $SU(5)$ multiplets and a right-handed $SO(5)$ multiplet:
\begin{eqnarray}
 \Psi_1 = \left( \begin{array}{c} \psi_1 \\ 0 \\ 0 \end{array}
 \right),~~~~  \Psi_2 = \left( \begin{array}{c} 0 \\ 0 \\ \psi_2 \end{array}
 \right),~~~~  \Psi_{HR} = \left( \begin{array}{c} \tilde{\psi}_{HR} \\
 \chi_{HR} \\ \psi_{HR} \end{array} \right),
\end{eqnarray}
with
\begin{eqnarray}
 \psi_A = -\tau^2 q_A  =  -\tau^2
 \left(
 \begin{array}{c}
 u_A \\
 d_A
 \end{array}
 \right),~~~~(A = 1, 2, HR),
\end{eqnarray}
where $\tau^2$ is the second Pauli matrix.
The transformations for these fields under $SU(5)$ tell us that $q_1$, $q_2$ and $q_{HR}$ are all $SU(2)_L$ doublets. Under $T$ parity,
$\Psi_1$, $\Psi_2$ and $\Psi_{HR}$ transform as
\begin{eqnarray}
\Psi_1 \longrightarrow -\Sigma_0 \Psi_2,~~~~~
\Psi_2 \longrightarrow -\Sigma_0 \Psi_1,~~~~~
\Psi_{HR} \longrightarrow -\Psi_{HR},
\end{eqnarray}
where $\Sigma_0$ is a $5 \times 5$ symmetric tensor defined as
\begin{eqnarray}
\Sigma_0 = \left(
 \begin{array}{ccccc}  & & 1_{2 \times 2} \\  & 1 & \\  1_{2 \times 2} & & \end{array}
 \right).
\end{eqnarray}
Thus, the $T$-parity eigenstates of the $SU(2)_L$ quark doublets $q_A~ (A = 1, 2, HR)$ are
\begin{eqnarray}
\label{T-eigenstates-1}
&& q_{SM} = \frac{q_1 - q_2}{\sqrt{2}},~~~~~~~~~~~~~~~~~~~~~~~(T-{\rm even}), \nonumber \\
&& q_{HL} = \frac{q_1 + q_2}{\sqrt{2}},~~~~~~~ q_{HR},~~~~~~~~~~(T-{\rm odd}).
\end{eqnarray}
$q_{SM}$ is the left-handed $SU(2)_L$ SM quark doublet, while $q_{HL}$ the left-handed $SU(2)_L$ mirror quark doublet. The right-handed $SU(2)_L$ mirror quark doublet is given by $q_{HR}$.

\par
The $T$-odd mirror quarks acquire masses via the following Lagrangian:
\begin{eqnarray}
 {\cal L}_{\rm mirror} = -\sum_{i,j=1}^3 \kappa f \delta_{ij}
 \Big(
 \bar{\Psi}_2^i \xi +
 \bar{\Psi}_1^i \Sigma_0 \Omega \xi^{\dag} \Omega
 \Big)
 \Psi_{HR}^j ~+ ~{\rm h.c.},
\end{eqnarray}
where $\Omega = {\rm diag}(1,1,-1,1,1)$, $\xi = e^{i \Pi/f}$, $i,j=1,...,3$ are flavor indices, and $\kappa$ is the mass coefficient of $T$-odd mirror quarks. Assuming a flavor independent coupling, the masses of the $T$-odd up- and down-type quarks at the ${\cal O}(v^2/f^2)$ are given by
\begin{eqnarray}
\label{q- mass}
m_{u_{i-}} = \sqrt{2}\kappa f\left( 1 - \frac{1}{8} \frac{v^2}{f^2} \right),~~~~~~ m_{d_{i-}} = \sqrt{2}\kappa f,
\end{eqnarray}
where $u_{i-} = u_-, c_-, t_-$ and $d_{i-} = d_-, s_-, b_-$ with $i$ running from 1 to 3.

\par
In order to cancel the large quadratic divergent correction to the Higgs boson mass induced by the top quark, a $T$-even top-quark partner $T_+$ is introduced. The implementation of $T$ parity then requires also a $T$-odd partner $T_-$. The Yukawa interaction for the top sector is given by
\begin{eqnarray}
\label{L-top}
&&{\cal L}_{\rm top} = -\frac{1}{2 \sqrt{2}} \lambda_1 f \epsilon_{ijk} \epsilon_{xy}
\left[ (\bar{Q}_1)_i (\Sigma)_{jx} (\Sigma)_{ky} -
(\bar{Q}_2 \Sigma_0)_i (\tilde{\Sigma})_{jx} (\tilde{\Sigma})_{ky}
\right] u_R \nonumber \\ &&~~~~~~~~~
-\lambda_2 f \left( \bar{U}_{L1} U_{R1} + \bar{U}_{L2} U_{R2} \right)~+ ~{\rm h.c.},
\end{eqnarray}
where $\Sigma = e^{2 i \Pi/f} \Sigma_0$, and $\tilde{\Sigma} = \Sigma_0 \Omega \Sigma^{\dag} \Omega \Sigma_0$ is the image of $\Sigma$ under $T$ parity. The $SU(5)$ multiplets $Q_1$ and $Q_2$ are defined as
\begin{eqnarray}
 Q_1 = \left( \begin{array}{c} \psi_1 \\ U_{L1} \\ 0
 \end{array}
 \right),~~~~~~~
 Q_2 =
 \left( \begin{array}{c} 0 \\ U_{L2} \\ \psi_2
 \end{array} \right),
\end{eqnarray}
which obey the same transformation laws under $T$ parity and $SU(5)$ as do $\Psi_1$ and $\Psi_2$. $U_{L1}$ and $U_{L2}$ are left-handed $SU(2)_L$ singlets. $U_{R1}$, $U_{R2}$ and $u_R$ are all right-handed $SU(2)_L$ singlets, and transform under $T$ parity as
\begin{eqnarray}
U_{R1} \leftrightarrow -U_{R2},~~~~~~~~ U_{R2} \leftrightarrow -U_{R1},~~~~~~~~ u_{R} \leftrightarrow u_{R}.
\end{eqnarray}
Then we obtain the following $T$-parity eigenstates
\begin{eqnarray}
 U_{L \pm} = \frac{U_{L1} \mp U_{L2}}{\sqrt{2}},~~~~~~ U_{R \pm} = \frac{U_{R1} \mp U_{R2}}{\sqrt{2}},~~~~~~ u_R,
\end{eqnarray}
in addition to those in (\ref{T-eigenstates-1}). From the top Yukawa Lagrangian (\ref{L-top}), we can get the mass eigenstates of the top sector. The $T$-odd eigenstates $U_{L-}$ and $U_{R-}$ do not mix with the $T$-odd mirror quarks, while the $T$-even eigenstates $U_{L+}$ and $U_{R+}$ mix with $u_{SM}$ and $u_R$, respectively, so that the mass eigenstates of the top quark $t$ and its heavy partners $T_{\pm}$ are given by
\begin{eqnarray}
&& \left(
\begin{array}{c}
t_L \\
\left(T_+\right)_L
\end{array}
\right)
=
\left(
\begin{array}{cr}
\cos\theta_L & -\sin\theta_L \\
\sin\theta_L & \cos\theta_L
\end{array}
\right)
\left(
\begin{array}{c}
u_{SM} \\
U_{L+}
\end{array}
\right),~~~~~~~
\left(T_-\right)_L = U_{L-},  \nonumber \\
&&
\left(
\begin{array}{c}
t_R \\
\left(T_+\right)_R
\end{array}
\right)
=
\left(
\begin{array}{cr}
\cos\theta_R & -\sin\theta_R \\
\sin\theta_R & \cos\theta_R
\end{array}
\right)
\left(
\begin{array}{c}
u_{R} \\
U_{R+}
\end{array}
\right),~~~~~~~
\left(T_-\right)_R = U_{R-}.
\end{eqnarray}
The masses of $T_+$ and $T_-$ can be expressed as
\begin{eqnarray}
&& m_{T_+}=
\frac{f}{v} \frac{m_t}{\sqrt{x_L (1 - x_L)}}
\left[
1 + \frac{v^2}{f^2} \left( \frac{1}{3} - x_L (1 - x_L) \right)
\right], \nonumber \\
&& m_{T_-}
=
\frac{f}{v} \frac{m_t}{\sqrt{x_L}}
\left[
1 + \frac{v^2}{f^2} \left( \frac{1}{3} - \frac{1}{2} x_L (1 - x_L) \right)
\right],
\end{eqnarray}
where
\begin{eqnarray}
m_{t}
=
v \sqrt{x_L (1 - x_L) (\lambda_1^2 + \lambda_2^2)}
\left[
1 + \frac{v^2}{f^2} \left( -\frac{1}{3} + \frac{1}{2} x_L (1 - x_L) \right)
\right],~~~~~~
x_L = \lambda_1^2/(\lambda_1^2 + \lambda_2^2),
\end{eqnarray}
and $\lambda_{1,2}$ are the Yukawa coupling constants of top sector.

\vskip 3mm
\section{Calculations}
\par
\begin{figure}
\begin{center}
\includegraphics[width=0.45\textwidth]{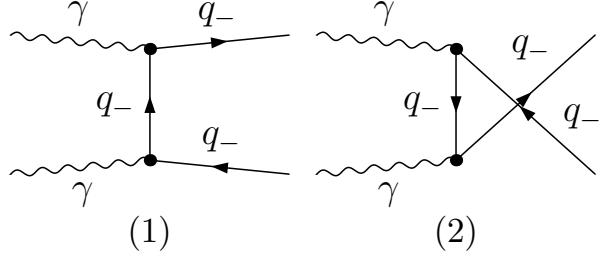}
\caption{ \label{LO-feyndiag} The tree-level Feynman diagrams for the $\gamma \gamma \to q_- \bar{q}_-$ process. }
\end{center}
\end{figure}

The $T$-odd mirror quark pair production at a photon-photon collider, denoted as $\gamma(p_1) + \gamma(p_2) \to q_-(p_3) + \bar{q}_-(p_4)$, is a pure electromagnetic process at the leading order. The tree-level Feynman diagrams are depicted in Fig.\ref{LO-feyndiag}. The LO differential cross section for this process is simply given by
\begin{eqnarray}
d{\sigma}_{LO}(\gamma\gamma \to q_-\bar{q}_-)
=
\frac{1}{4}\frac{(2\pi)^4N_{c}}{4 |\vec{p}_{1}| \sqrt{s}}
\sum_{\rm spin} \left|{\cal M}_t + {\cal M}_u\right|^2 d\Phi_2,
\end{eqnarray}
where $N_c=3$, $\sqrt{s}$ is the center-of-mass energy of the photon-photon collision, the summation is taken over the spins of the initial and final states, $d \Phi_2$ is the two-body phase-space element defined as
\begin{eqnarray}
d\Phi_2=\delta^{(4)} \left( p_1+p_2-p_3-p_4 \right) \prod_{i=3}^4
\frac{d^3 \vec{p}_i}{(2 \pi)^3 2 E_i},
\end{eqnarray}
and ${\cal M}_t$, ${\cal M}_u$ are the amplitudes for the $t$- and $u$-channel Feynman diagrams, respectively, expressed as
\begin{eqnarray}
{\cal M}_t =
-i Q_{q_-}^2 \bar{u}(p_3) \rlap/\epsilon(p_1) \frac{1}{\rlap/p_3 - \rlap/p_1 - m_{q_-}} \rlap/\epsilon(p_2) v(p_4),~~~~~~~
{\cal M}_u = {\cal M}_t \Big|_{p_1 \leftrightarrow p_2}.
\end{eqnarray}
Since the third generation of $T$-odd mirror quarks, $b_-$ and $t_-$, can be identified in experiment, we only consider the $T$-odd mirror quark pair production of the first two generations in $\gamma \gamma$ collision. Both the total and differential cross sections for the $\gamma \gamma \to q_- \bar{q}_-$ process presented in this paper are summed over the $u_-\bar{u}_-$, $d_-\bar{d}_-$, $c_-\bar{c}_-$ and $s_-\bar{s}_-$ production processes.

\par
The NLO QCD corrections to the $\gamma\gamma \to q_-\bar{q}_-$ process include virtual loop and real emission corrections, which contain ultraviolet (UV) and infrared (IR) singularities. We adopt the dimensional scheme for renormalization, in which the dimensions of both the spinor and space-time manifolds are extend to $D = 4 - 2 \epsilon$, to regularize these divergences. Since the physical observables are UV- and IR-finite, all the UV and IR divergences should be canceled. The virtual loop correction contains both the UV and IR singularities. In the on-mass-shell renormalization scheme, the $T$-odd mirror quark wave function and mass renormalization constants are expressed as
\begin{eqnarray}
&& \delta Z^{L,R}_{q_-} = -\frac{\alpha_s (\mu_r)}{3\pi} \left[ \Delta_{UV} + 2 \Delta_{IR} + 4 + 3 \ln \left( \frac{\mu_r^2}{m_{q_-}^2} \right) \right], \nonumber \\
&& \frac{\delta m_{q_-}}{m_{q_-}}=-\frac{\alpha_s(\mu_r)}{3\pi} \left\{ 3\left[ \Delta_{UV} + \ln \left( \frac{\mu_r^2}{m_{q_-}^2} \right) \right] + 4 \right\},
\end{eqnarray}
where $\Delta_{UV}=\frac{1}{\epsilon_{UV}}-\gamma_E + \ln (4\pi)$ and $\Delta_{IR}=\frac{1}{\epsilon_{IR}}-\gamma_E + \ln (4\pi)$ are the UV and IR regulators, respectively. The UV divergence associated with these renormalization constants can cancel exactly those arising from the loop integrals. Then the virtual correction is UV-finite after performing the renormalization procedure.

\par
Although the renormalized virtual correction is UV-finite, it still contains the IR-soft singularity. According to the Kinoshita-Lee-Nauenberg theorem \cite{KLN-1,KLN-2}, the IR-soft singularity of the renormalized virtual correction can be canceled by that of the real gluon emission correction. We denote the real gluon emission process for the $T$-odd mirror quark pair production at a $\gamma \gamma$ collider as $\gamma(p_1) + \gamma(p_2) \to q_-(p_3) + \bar{q}_-(p_4) + g(p_5)$. In our NLO calculations for the $\gamma \gamma \to q_- \bar{q}_-$ process there exists QCD soft IR-singularity, but no QCD collinear IR-singularity. In order to manipulate the IR-soft singularity, we employ the phase-space slicing method proposed by B. W. Harris and J. F. Owens\cite{tcpss}. A soft cutoff $\delta_s$ is introduced to separate the phase space of the $\gamma \gamma \to q_- \bar{q}_- g$ process into two regions, the soft gluon region ($E_5\leq \frac{1}{2}\delta_s \sqrt{s}$) and the hard gluon region ($E_5> \frac{1}{2}\delta_s \sqrt{s}$). The IR-soft singularity is located at the soft gluon region, while the phase-space integration over the hard gluon region is finite and can be evaluated in the four dimensions by using the Monte Carlo technique.

\par
Finally, the total cross section for the $T$-odd mirror quark pair production at a photon-photon collider at the QCD NLO can be expressed as
\begin{eqnarray}
\sigma_{NLO} = \sigma_{LO} + \Delta\sigma_{NLO} = \sigma_{LO} + \Big[ \sigma^{V} + \sigma^R_{S} \Big] + \sigma^{R}_{H},
\end{eqnarray}
where $\sigma^{V}$, $\sigma^{R}_{S}$ and $\sigma^{R}_{H}$ are the renormalized virtual correction, real soft gluon emission correction and real hard gluon emission correction, respectively. As mentioned above, both $ \Big[ \sigma^{V} + \sigma^R_{S} \Big]$ and $\sigma^{R}_{H}$ are finite, and therefore can be calculated numerically.

\par
The $T$-odd mirror quarks are unstable particles, thus we should consider their decay channels in studying the kinematic distributions of the final products of the $T$-odd mirror quark pair production. The main decay modes of the $T$-odd mirror quark $q_-$ are $q_- \to W_H q^{\prime}$, $q_- \to Z_H q$ and $q_- \to A_H q$, i.e., $\Gamma_{q_-}^{\rm tot} \simeq \Gamma(q_- \to W_H q^{\prime}) + \Gamma(q_- \to Z_H q) + \Gamma(q_- \to A_H q)$, with the LO decay widths expressed as \cite{Du}
\begin{eqnarray}\label{DecayWidth}
\Gamma_{LO} (q_{i-} \to A_H q_j)&=&
\frac{\left|(V_{Hq})_{ij} \right|^2}{32 \pi} \left( \frac{g}{2} \sin\theta_H + I_{q_j} \frac{g^{\prime}}{5} \cos\theta_H \right)^2
\frac{m_{q_{i-}}^3}{m_{A_H}^2}
\left( 1- \frac{m_{A_H}^2}{m_{q_{i-}}^2} \right)^2
\left( 1+ \frac{2 m_{A_H}^2}{m_{q_{i-}}^2}\right), \nonumber \\
\Gamma_{LO} (q_{i-} \to Z_H q_j)&=&
\frac{\left|(V_{Hq})_{ij} \right|^2}{32 \pi} \left( \frac{g}{2} \cos\theta_H - I_{q_j}  \frac{g^{\prime}}{5} \sin\theta_H \right)^2
\frac{m_{q_{i-}}^3}{m_{Z_H}^2}
\left( 1- \frac{m_{Z_H}^2}{m_{q_{i-}}^2} \right)^2
\left( 1+ \frac{2 m_{Z_H}^2}{m_{q_{i-}}^2}\right), \nonumber \\
\Gamma_{LO} (q_{i-} \to W_H q^{\prime}_j)&=&
\frac{\left|(V_{Hq^{\prime}})_{ij} \right|^2}{64 \pi} g^2
\frac{m_{q_{i-}}^3}{m_{A_H}^2}
\left( 1- \frac{m_{A_H}^2}{m_{q_{i-}}^2} \right)^2
\left( 1+ \frac{2 m_{A_H}^2}{m_{q_{i-}}^2}\right), ~~~~(i, j = 1, 2),
\end{eqnarray}
where the the first two generation quarks are taken to be massless and
\begin{eqnarray}
q^{\prime}_j= \left\{
\begin{array}{l}
d,s,~~~~~(for~q_{i-} = u_{-},c_{-}) \\
u,c,~~~~~(for~q_{i-} = d_{-},s_{-})
\end{array}
\right., ~~~~~~~
 I_{q_j} = \left\{
\begin{array}{l}
+\frac{1}{2},~~~~~(q_j = u,c) \\
-\frac{1}{2},~~~~~(q_j = d,s),
\end{array}
\right.
\end{eqnarray}
and $V_{Hu}$, $V_{Hd}$ are CKM-like unitary mixing matrices satisfying $V_{Hu}^{\dag} V_{Hd} = V_{CKM}$. We neglect the mixing between the first two generations and the third generation. At the QCD NLO, the partial decay width for the $q_{i-} \to V_H q^{\prime}_j$ decay mode has the form as
\begin{eqnarray}\label{Width-1}
\Gamma_{NLO} (q_{i-} \to V_H q^{\prime}_j)=
\Gamma_{LO} (q_{i-} \to V_H q^{\prime}_j) \times \left[ 1 + \frac{2 \alpha_s}{3 \pi} \left( \frac{2 \pi^2}{3} - \frac{5}{2} \right) + {\cal O} \left( \frac{\alpha_s}{\pi} \frac{m_{V_H}^2}{m_{q_{i-}}^2} \right) \right].
\end{eqnarray}
We evaluate the terms of ${\cal O} \left( \frac{\alpha_s}{\pi} \frac{m_{V_H}^2}{m_{q_-}^2} \right)$ ($V_H = A_H, Z_H, W_H$ and $q_- = u_-, d_-, c_-, s_-$), and find that their contributions to the branch ratios are less than $0.01\%$ in the parameter space adopted in this work, and therefore can be neglected.

\par
We use the FeynArts-3.4 package developed by us to generate Feynman diagrams and their corresponding amplitudes for the $\gamma\gamma \to q_-\bar{q}_-$ process in the LHT at both the LO and QCD NLO, and employ the FormCalc-5.4 program for algebraic manipulation. The IR singularities of loop integrals are isolated analytically by using our developed-in-house programs, and the IR-finite parts are calculated numerically based on the LoopTools-2.4 package. The analytical expressions for the IR-singular parts of loop integrals and the formulas for evaluating the IR-finite $N$-point ($N \leq 4$) integrals are given in Refs.\cite{Stefan,OneTwoThree,Four}, respectively.

\vskip 3mm
\section{Numerical results and discussions}
\par
In this section we provide and discuss the numerical results for the $T$-odd mirror quark pair production at a $\gamma \gamma$ collider in the LHT up to the QCD NLO. The SM electroweak input parameters for our calculations are taken as: $\alpha_{\rm ew}^{-1}=137.036$, $m_W = 80.385~{\rm GeV}$ and $m_Z = 91.1876~{\rm GeV}$ \cite{PDG2012}. The $u$-, $d$-, $c$- and $s$-quarks are treated as massless particles, and the CKM matrix is set to be the unit matrix, i.e., $V_{CKM} = I$. Considering the latest results from the $8~{\rm TeV}$ run at the LHC, the constraints from Higgs couplings are by now competing with electroweak precision tests and both exclude $f$ up to $694~{\rm GeV}$ or $560~{\rm GeV}$ depending on the implementation of the down-type Yukawa sector \cite{LHTlimits}.
In our numerical calculations, we constrain the global symmetry breaking scale of the LHT in the range of $f \ge 700~{\rm GeV}$. The $T$-odd mirror quark mass coefficient $\kappa$ is fixed to be $1$ in default unless stated otherwise, and the two CKM-like matrices are take as $V_{Hu} = I$ and $V_{Hd} = V_{CKM}$. In the NLO QCD calculations for the $\gamma \gamma \to q_- \bar{q}_-$ process, we set the QCD renormalization scale being $\mu=\mu_r= m_{q_-}$. In Table \ref{Mass-H} we present the masses of heavy particles in the LHT for some typical values of the LHT global symmetry breaking scale $f$.
\begin{table}
\begin{center}
\begin{tabular}{c|c|c|c|c|c|c}
  \hline
    $f$    & $m_{u_-}$  & $m_{d_-}$  & $m_{W_H}\approx m_{Z_H}$  & $m_{A_H}$ & $m_{T_+}$ & $m_{T_-}$ \\
   $~~({\rm GeV})~~$ & $~~({\rm GeV})~~$ & $~~({\rm GeV})~~$
   &  $~~({\rm GeV})~~$  &  $~~({\rm GeV})~~$ & $~~({\rm GeV})~~$ & $~~({\rm GeV})~~$ \\
  \hline
     700  & 974.7   & 989.9    & 442.1  &   99.2 & 996.6 & 715.5 \\
     800  & 1118.0  & 1131.4   & 507.1  &  115.6 & 1136.3 & 812.9 \\
     900  & 1260.9  & 1272.8   & 571.9  &  131.8 & 1276.2 & 910.8 \\
     1000 & 1403.5  & 1414.2   & 636.6  &  147.8 & 1416.4 & 1009.1 \\
  \hline
\end{tabular}
\end{center}
\begin{center}
\caption{\label{Mass-H} The masses of relevant heavy new particles in the LHT for some typical values of $f$ with $\kappa =1$ and $x_L = 1/2$. ($x_L$ is only used in calculating the masses of $T_+$ and $T_-$.) }
\end{center}
\end{table}

\par
Since the summation of all the QCD NLO contributions should be numerically finite, we verify the correctness of our calculation by means of confirming the cancellations of the UV and IR divergences. To isolate the soft IR divergences, we introduce an arbitrary cutoff $\delta_s$ to separate the phase space. The total NLO QCD corrected cross section ($\sigma_{NLO}$) for the $\gamma \gamma \to q_- \bar q_-$  process is obtained by summing up the two-body and three-body contribution parts ($\sigma_{LO}+\Delta \sigma^{(2)} = \sigma_{LO} + \sigma^V + \sigma_S^R$ and $\Delta \sigma^{(3)} = \sigma_H^R$). The $\sigma_{NLO}$ should be independent of $\delta_s$ \cite{tcpss}.
We check the independence of the NLO QCD corrected integrated cross section on the cutoff $\delta_s$ within the statistical errors by varying the cutoff $\delta_s$ in the range of $\left[ 1\times 10^{-6},~1\times 10^{-4}\right]$ with $\sqrt{s}=3~{\rm TeV}$, $\kappa=1$ and $f=700~{\rm GeV}$. This is also an indirect verification for the correctness of our calculation. In further numerical calculations, we fix $\delta_s = 1\times 10^{-4}$.

\par
\begin{figure}
\begin{center}
\includegraphics[scale=0.6]{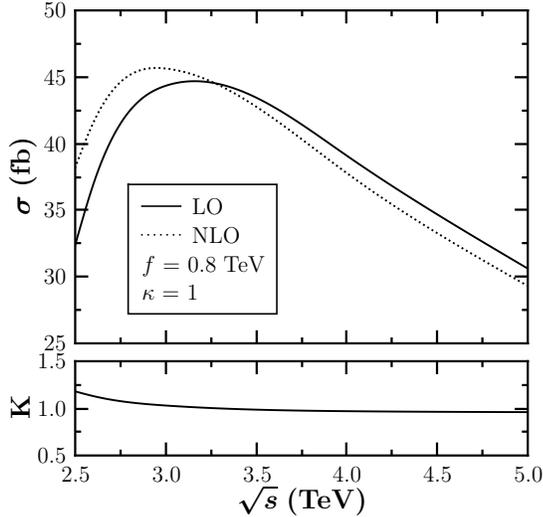}
\caption{ \label{fig3} The LO, NLO QCD corrected cross sections and the corresponding $K$-factor for $\gamma\gamma \to q_- \bar q_-$ as functions of the colliding energy $\sqrt{s}$ with $f=800~{\rm GeV}$. }
\end{center}
\end{figure}

To study the dependence of cross sections on the colliding energy $\sqrt{s}$, we plot the LO, NLO QCD corrected cross sections ($\sigma_{LO}$, $\sigma_{NLO}$) and the corresponding $K$-factor defined as $K = \frac{\sigma_{NLO}}{\sigma_{LO}}$ for the $\gamma \gamma \to q_- \bar q_-$ process as the functions of $\sqrt{s}$ with $f=800~{\rm GeV}$ in Fig.\ref{fig3}. Both the LO and NLO QCD corrected cross sections obviously go up with the increment of colliding energy in the range near the $q_-\bar q_-$ production threshold, while decrease rapidly when $\sqrt{s}$ goes up beyond $3200~{\rm GeV}$ as shown in the upper plot of Fig.\ref{fig3}. And we can see from the lower plot of Fig.\ref{fig3}, the $K$-factor has a relative large value in the vicinity of the $q_-\bar q_-$ threshold.

\par
\begin{figure}[htbp]
\begin{center}
\includegraphics[scale=0.6]{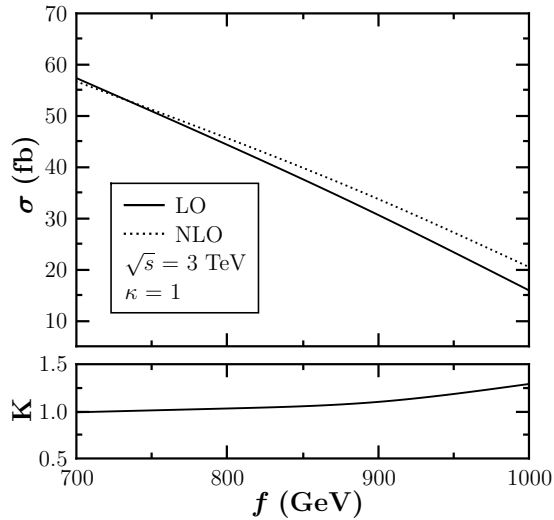}
\hspace{0in}%
\caption{\label{fig4} The LO, NLO QCD corrected integrated cross sections and the corresponding $K$-factor for the $\gamma \gamma \to q_- \bar q_-$ process as functions of the global symmetry breaking scale $f$ at a $\sqrt{s}=3~{\rm TeV}$ photon-photon collider. }
\end{center}
\end{figure}

The LO and NLO QCD corrected integrated cross sections at a $\sqrt{s}=3~{\rm TeV}$ photon-photon collider and the corresponding $K$-factors as functions of the global symmetry breaking scale $f$ are plotted in Fig.\ref{fig4}, separately. Since the masses of final $q_-$ and $\bar q_-$ become heavier with the increment of $f$ and make the phase space getting smaller, the LO and NLO QCD corrected total cross sections for the $\gamma \gamma \to q_- \bar q_-$ process would decrease with the increment of $f$ as demonstrated in Fig.\ref{fig4}. The $K$-factor in the lower plot of Fig.\ref{fig4} gradually increases with the increment of $f$ due to the fact that the threshold value ($2 m_{q_-}$) moves towards the colliding energy with the increment of $f$. From Fig.\ref{fig4} we read out some numerical results for the $\gamma \gamma \to q_- \bar q_-$ process at a $\sqrt{s}=3~{\rm TeV}$ photon-photon collider for some typical values of $f$, and listed them in Table \ref{tab2}.
\begin{table}
\begin{center}
\begin{tabular}{c|c|c|c}
\hline
$f$ ({\rm GeV}) & $\sigma_{LO}$ ({\rm fb}) & $\sigma_{NLO}$ ({\rm fb})&$K$ \\
\hline
  700   & 57.3691(1)   & 56.708(3)  & 0.99      \\
  800   & 44.34106(9)  & 45.6365(9) & 1.03      \\
  900   & 30.65976(9)  & 33.7591(4) & 1.10      \\
  1000  & 15.92042(1)  & 20.48004(6)& 1.29      \\
\hline
\end{tabular}
\end{center}
\begin{center}
\begin{minipage}{15cm}
\caption{\label{tab2}
The LO, NLO QCD corrected cross sections and the corresponding $K$-factor for the $\gamma\gamma \to q_- \bar q_-$ process for some typical values of $f$ at a $\sqrt{s}=3~{\rm TeV}$ photon-photon collider.  }
\end{minipage}
\end{center}
\end{table}

\par
\begin{figure}[htbp]
\begin{center}
\includegraphics[scale=0.45]{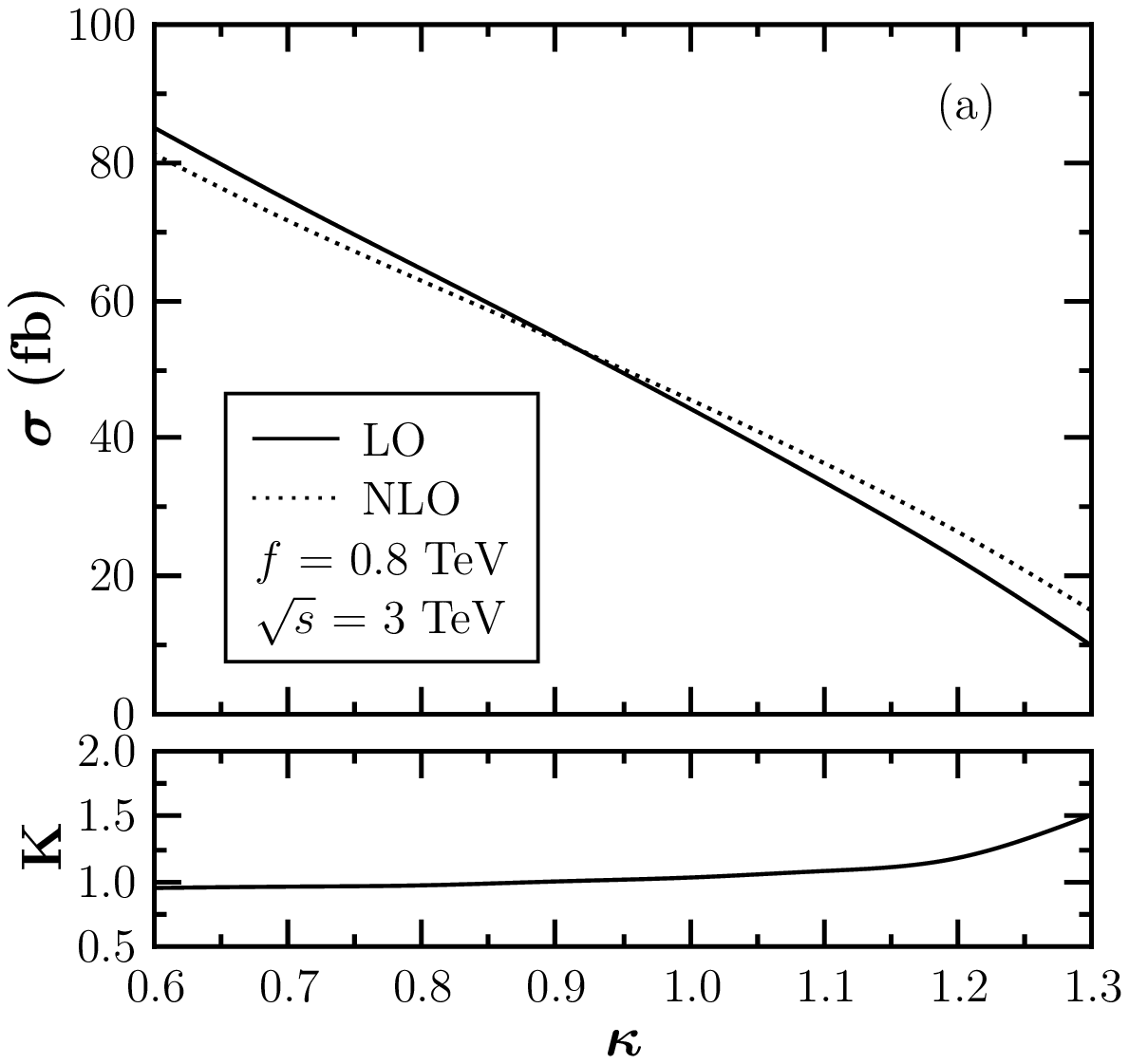}
\includegraphics[scale=0.45]{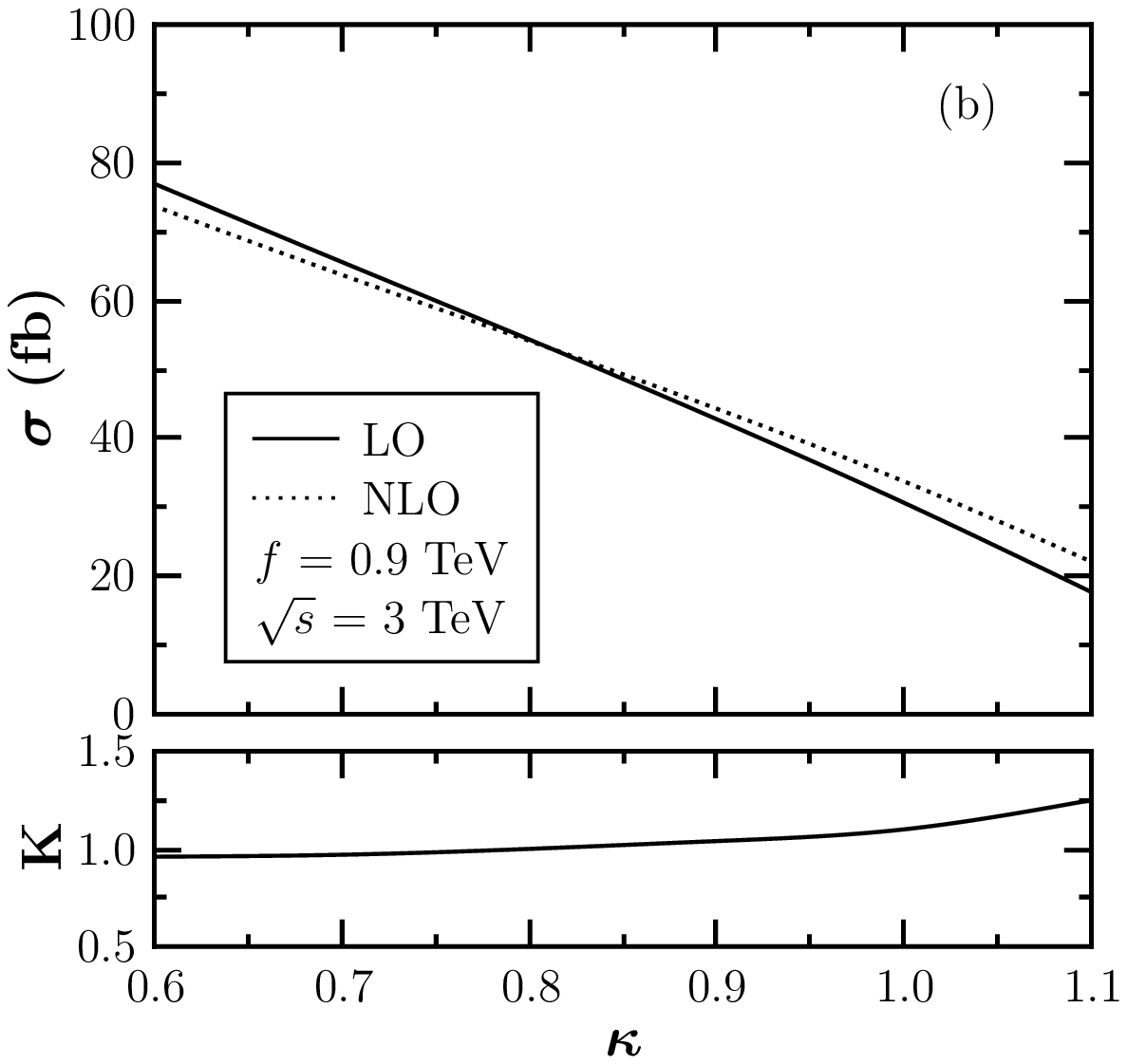}
\includegraphics[scale=0.45]{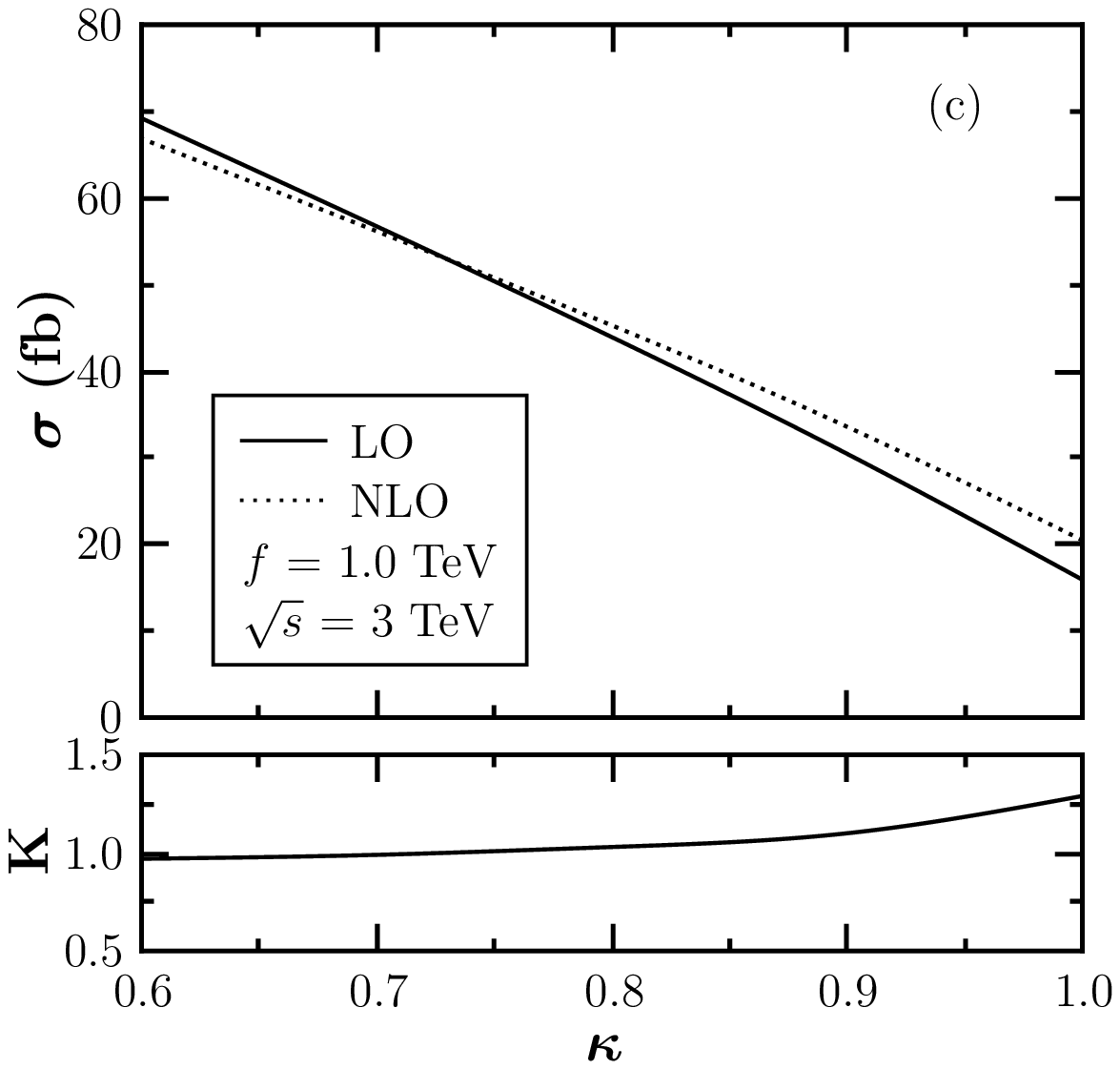}
\hspace{0in}%
\caption{\label{fig5} The LO, NLO QCD corrected cross sections and the corresponding $K$-factors for the $\gamma\gamma \to q_- \bar q_-$ process as functions of the $T$-odd mirror quark mass coefficient $\kappa$ at the $\sqrt{s}=3~{\rm TeV}$ photon-photon collider. (a) $f=0.8~{\rm TeV}$. (b) $f=0.9~{\rm TeV}$. (c) $f=1.0~{\rm TeV}$.  }
\end{center}
\end{figure}

The LO, NLO QCD corrected integrated cross sections and the corresponding $K$-factors versus the LHT $T$-odd quark mass coefficient $\kappa$ at the $\sqrt{s}=3~{\rm TeV}$ photon-photon collider with $f = 0.8$, $0.9$ and $1.0~ {\rm TeV}$ are plotted in Figs.\ref{fig5}(a-c), separately. As shown in  these figures both the LO and NLO QCD corrected cross sections decrease with the increment of $\kappa$, while the $K$-factor increases with the increase of $\kappa$. That is because the threshold value approaches the colliding energy when $\kappa$ goes up. One can read out from the figures that for $f = 0.8~{\rm TeV}$ the corresponding K-factor varies from $0.95$ to $1.51$ with $\kappa$ going up from $0.6$ to $1.3$, for $f = 0.9~{\rm TeV}$ the $K$-factor increases from $0.96$ to $1.25$ with $\kappa$ running from $0.6$ to $1.1$, while for $f = 1.0~{\rm TeV}$ the $K$-factor goes up from $0.97$ to $1.29$ with $\kappa$ increasing from $0.6$ to $1.0$.

\par
\begin{figure}[htbp]
\begin{center}
\includegraphics[scale=0.6]{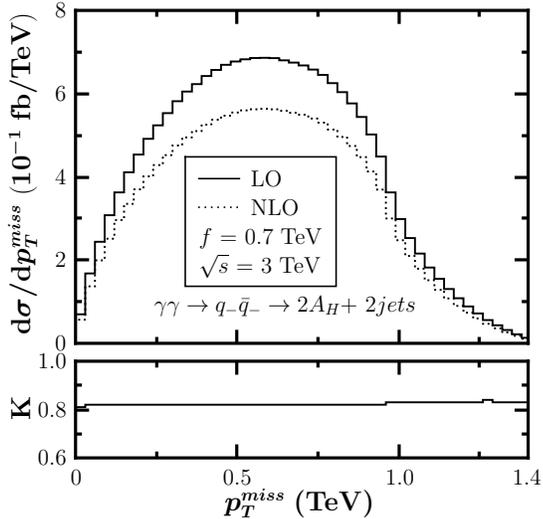}
\hspace{0in}%
\caption{\label{fig6} The LO, NLO QCD corrected distributions of the missing transverse momentum $(p_{T}^{miss})$, and the corresponding $K$-factor for the $\gamma \gamma \to q_-\bar q_- \to 2 A_H + 2 jets$ process at a $\sqrt{s}=3~{\rm TeV}$ photon-photon collider. }
\end{center}
\end{figure}

In the following, we investigate the kinematic distributions of final particles after the on-shell $T$-odd mirror quark decays ($q_-[\bar{q}_-] \to A_H q[\bar{q}]$), adopting the narrow width approximation. The $q_-[\bar{q}_-]$ decay products involve a $q[\bar q]$-jet and missing energy of the lightest neutral stable particle $A_H$. Its SM background should be mainly from the $\gamma\gamma \to q \bar q Z \to q \bar q \nu \bar \nu$ process, where $q\bar q = u\bar u, c\bar c,
d\bar d, s\bar s$. According to Eq.(\ref{Width-1}) one can obtain the $q_- \to A_H q$ decay branch ratio up to the QCD NLO. After the decays of the $T$-odd mirror quarks, we may encounter the event with several jets in the analyses. We adopt the Cambridge/Aachen (C/A) jet algorithm \cite{Cacciari:2008gp,jet-algorithm} provided by the \texttt{Fastjet} package\cite{Cacciari:2011ma} with the distance parameter $R=0.4$ to merge the proto-jets. The recombined $i$-th and $j$-th jets gain its four-momentum by $p_{\mu}^{ij}=p_{\mu}^i+p_{\mu}^j$. After the jet merging procedure, theoretically we may meet three kind of events, i.e., 1-jet, 2-jet and 3-jet events. In the following analyses, we collect the ``2-jet" events by adopting the exclusive 2-jet event selection criterion to collect the signal and background events:
\begin{enumerate}
\item[(I)]  For theoretical 2-jets events, we accept the event with both two jets satisfying the condition of $p_T^j~>~ 20~{\rm GeV}$.
\item[(II)] For theoretical 3-jets events, we accept the event with only two hardest jets satisfying the constraints of $p_T^j~>~ 20~{\rm GeV}$ and the remained jet satistying $p_T^j~<~ 20~{\rm GeV}$.
\end{enumerate}
In Fig.\ref{fig6} we plot the LO and NLO QCD corrected distributions of the final missing transverse momentum $(\slashed{p}_{T}^{miss})$ for the $\gamma \gamma \to q_-\bar q_- \to 2 A_H + 2 jets$ process at a $\sqrt{s}=3~{\rm TeV}$ photon-photon collider, and the corresponding $K$-factor defined as $K(p_{T}^{miss})\equiv\frac{d\sigma_{NLO}/dp_{T}^{miss}}{d\sigma_{LO}/dp_{T}^{miss}}$. There we take $f=700~{\rm GeV}$ and get $m_{A_H} = 99.2~{\rm GeV}$. Both the LO and NLO differential cross sections reach their maxima in the vicinity of $p_{T}^{miss} \sim 580~{\rm GeV}$, and the corresponding $K$-factor varies from $0.81$ to $0.84$ when $p_{T}^{miss}$ goes up from $0$ to $1400~{\rm GeV}$ as displayed in the figure.

\par
\begin{figure}[htbp]
\begin{center}
\includegraphics[scale=0.5]{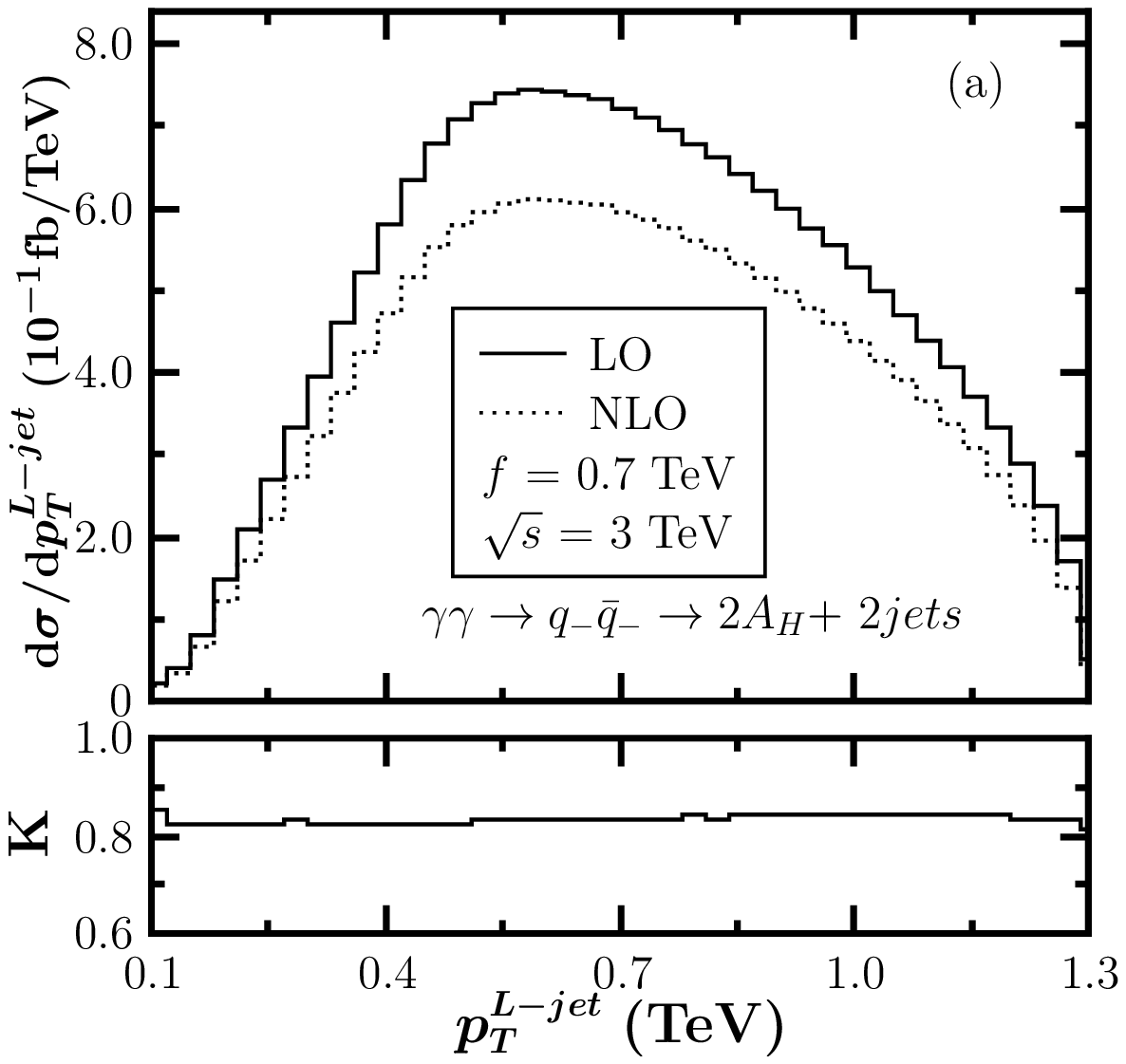}
\includegraphics[scale=0.5]{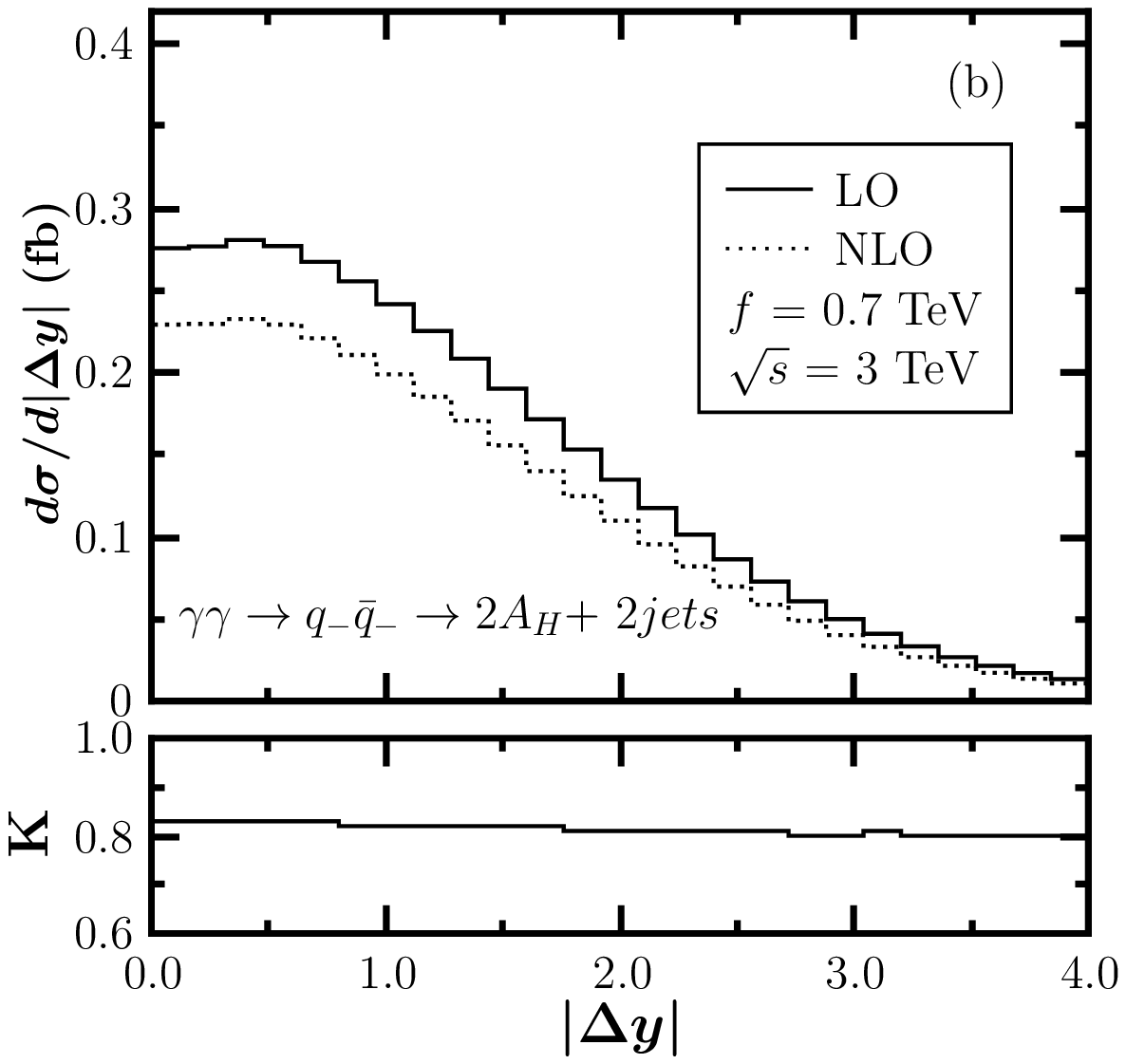}
\hspace{0in}%
\caption{\label{fig7} (a) The LO, NLO QCD corrected transverse momentum distributions of the leading jet and the corresponding $K$-factor for the $\gamma \gamma \to q_-\bar q_- \to 2 A_H + 2 jets$ process at the $\sqrt{s} = 3~{\rm TeV}$ photon-photon collider. (b) The LO, NLO QCD corrected distributions and the corresponding $K$-factor as functions of the rapidity separation of the final leading jet and next-to-leading jet $|\Delta y|\equiv|y_{L-jet}-y_{NL-jet}|$ at the $\sqrt{s} = 3~{\rm TeV}$ photon-photon collider. }
\end{center}
\end{figure}

In a ``2-jet" event, we name the jets with the largest energy and the second largest energy as the leading jet and next-to-leading jet, respectively. In Fig.\ref{fig7}(a) we present the LO and NLO QCD corrected transverse momentum distributions of the leading jet at the $\sqrt{s}=3~{\rm TeV}$ photon-photon collider and the corresponding $K$-factor defined as $\left[ K(p_{T}^{L-jet})\equiv\frac{d\sigma_{NLO}/dp_{T}^{L-jet}}{d\sigma_{LO}/dp_{T}^{L-jet}}\right]$. Here we take $f=700~{\rm GeV}$, and then get $m_{A_H} = 99.2~{\rm GeV}$. The peaks for the LO and QCD NLO curves are located in the vicinity of $p_T^{L-jet}\sim 600~{\rm GeV}$, and the $K$-factor varies in the range of $[0.80,~0.83]$ when $p_T^{L-jet}$ goes up from $100~{\rm GeV}$ to $1300~{\rm GeV}$ as shown in the figure.

\par
In Fig.\ref{fig7}(b) we plot the LO and NLO QCD corrected distributions of the rapidity separation of the final leading-jet and next-to-leading jet ($|\Delta y|\equiv |y_{L-jet}-y_{NL-jet}|$), and the corresponding $K$-factor ($K(|\Delta y|)\equiv\frac{d\sigma_{NLO}/d|\Delta y|}{d\sigma_{LO}/d|\Delta y|}$). We see that most of the $\gamma \gamma \to q_-\bar q_- \to 2 A_H + 2 jets$ events are concentrated in the low $|\Delta y|$ region and the $K$-factor varies between $0.80$ and $0.83$ with $|\Delta y|$ in the range of $[0,~4]$. All the distributions in Fig.\ref{fig6} and Figs.\ref{fig7}(a,b) show that the $K$-factor is not very sensitive to the kinematic variables, such as the transverse momentum, jet rapidity separation etc. But it would be necessary to calculate the complete NLO QCD corrections to get reliable kinematic distributions in the precision measurement.

\par
\begin{figure}[htbp]
\begin{center}
\includegraphics[scale=0.6]{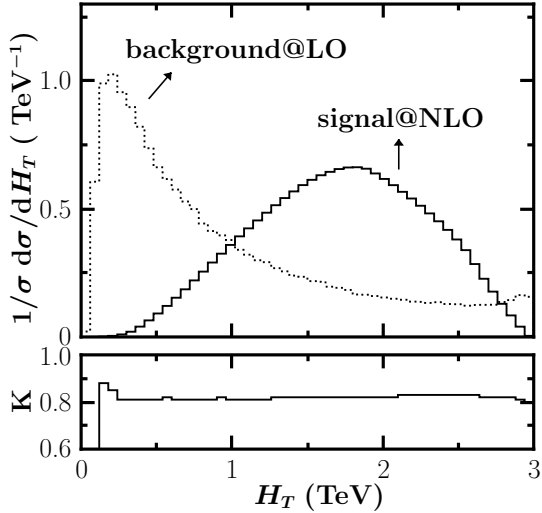}
\hspace{0in}%
\caption{\label{fig8} The normalized $H_T$ distributions for the signal process $\gamma \gamma \to q_- \bar q_- \to 2 A_H + 2 jets$ up to the QCD NLO with $f = 700~{\rm GeV}$ and the SM background process $\gamma \gamma \to q \bar q Z \to q \bar q \nu \bar \nu$ at the LO at a $3~TeV$ photon-photon collider. The corresponding QCD $K$-factors for the signal process are plotted in the lower panel.}
\end{center}
\end{figure}

For the $\gamma \gamma \to q_-\bar q_- \to 2 A_H + 2 jets$ signal process at a photon-photon collider, the main SM background comes from the $\gamma \gamma \to q \bar q Z \to q \bar q \nu \bar {\nu}$ process with two resolved jets. We define parameter $H_T$ as $H_T = \sum_i  | \vec{p}_T(i)|$ which is the scalar sum of the transverse momenta of all the final jets for both signal and background events. By adopting the exclusive 2-jet event selection scheme mentioned above, we present the normalized total QCD corrected $H_T$ distributions for the signal, and the LO distribution for its SM background in the upper panel of Fig.\ref{fig8}. There the distributions are normalized by the corresponding total cross sections. We can see that the SM background events tend to be concentrated in the low $H_T$ region with a peak in the vicinity of $H_T \sim 220~{\rm GeV}$ and then its event number declines rapidly. While the total QCD corrected $H_T$ distribution for the $q_-$-pair production signature has flatter peak in the vicinity of $H_T \sim 1800~{\rm GeV}$ and descends slowly as illustrated in Fig.\ref{fig8}. That indicates if we take proper lower limit on $H_T$ parameter, the background from the $\gamma \gamma \to q \bar q Z \rightarrow q \bar q \nu \bar {\nu}$ process can be significantly suppressed. In the lower figure of Fig.\ref{fig8} we show the corresponding $K$-factors of the $H_T$ distribution for the signal process, where we can see that the $K$-factor for the signature $H_T$ distribution varies in the range of $0.80 - 0.88$ with the increment of $H_T$ from $120~{\rm GeV}$ to $3~{\rm TeV}$.

\vskip 3mm
\section{Summary}
\par
In this paper we present the precision calculations of the $T$-odd mirror quark pair production including subsequent weak decays at a photon-photon collider up to the QCD NLO in the littlest Higgs model with $T$-parity.
The relationship between the LO, NLO QCD corrected integrated cross sections and the colliding energy $\sqrt{s}$ is investigated, and the LO and NLO QCD corrected kinematic distributions of final decay products are presented. We find that the $K$-factor is dependent on the phase space region, and increases significantly when the colliding energy approaches the $q_-$-pair production threshold. We see that the $K$-factor for the integrated cross section varies in the ranges of $0.99 \sim 1.29$ with $f$ in the range of $f \in [700,~ 1000]~{\rm GeV}$ at a $\sqrt{s} = 3~{\rm TeV}$ photon-photon collider. We conclude that NLO QCD corrections make relevant quantum impact on the $\gamma \gamma \to q_-\bar q_- \to 2 A_H + 2 jets$ processes, and should be included in any reliable analysis. We compare the $H_T$ distributions for the $T$-odd quark pair production signal and the main SM background, and conclude that they are remarkably different, and $\gamma \gamma \to q_-\bar q_- \to 2 A_H + 2 jets$ signal events can be discriminated from the possible $\gamma \gamma \to q \bar q Z \to q \bar q \nu \bar {\nu}$ background by taking proper cut on the $H_T$.

\vskip 3mm
\par
\section{Acknowledgments}
This work was supported in part by the National Natural Science Foundation of China (Grants. No.11275190, No.11375008, No.11375171).


\vskip 5mm
\begin{thebibliography}{99}

\bibitem{Chatrchyan:2012ufa}
  S. Chatrchyan {\it et al.}  [CMS Collaboration],
  Phys. Lett. {\bf B716} (2012) 30.

\bibitem{Aad:2012tfa}
  G. Aad {\it et al.}  [ATLAS Collaboration],
  Phys. Lett. {\bf B716} (2012) 1.

\bibitem{LTM-1}
  N. Arkani-Hamed, A.G. Cohen, and H. Georgi, Phys. Lett. {\bf B513} (2001) 232.

\bibitem{LTM-2}
  M. Schmaltz and D. Tucker-Smith, Annu. Rev. Nucl. Part. Sci. {\bf 55} (2005) 229.

\bibitem{LTM-3}
  M. Perelstein, Prog. Part. Nucl. Phys. {\bf 58} (2007) 247, and references therein.

\bibitem{T-parity-1}
   H.-C. Cheng, I. Low, JHEP {\bf 09} (2003) 051.

\bibitem{T-parity-2}
   H.-C. Cheng, I. Low, JHEP {\bf 08} (2004) 061.

\bibitem{Low:2004xc}
  I. Low, JHEP {\bf 10} (2004) 067.

\bibitem{Hubisz:2004ft}
  J. Hubisz and P. Meade, Phys. Rev. {\bf D71} (2005) 035016.

\bibitem{Hubisz:2005tx}
  J. Hubisz, P. Meade, A. Noble and M. Perelstein, JHEP {\bf 01} (2006) 135.

\bibitem{Barbieri:2000gf}
  R. Barbieri and A. Strumia, ``The `LEP paradox'", arXiv:hep-ph/0007265.

\bibitem{Cheng:2003ju}
  H.C. Cheng and I. Low, JHEP {\bf 09} (2003) 051; {\bf 08} (2004) 061.

\bibitem{Braun}
  H. Braun, \textit{et al}., CLIC-NOTE-764, [CLIC Study Team Collaboration], CLIC 2008 parameters, http://www.clic-study.org.

\bibitem{Abramowicz}
  H. Abramowicz, \textit{et al}., ``Physics at the CLIC $e^+e^-$ Linear Collider -- Input to the Snowmass process 2013'', arXiv:1307.5288.

\bibitem{Ginzburg-1}
  I.F. Ginzburg, G.L. Kotkin, S.L. Panfil, V.G. Serbo and V.I. Telnov, Nucl. Instrum. Meth. {\bf 219} (1984) 5.

\bibitem{Ginzburg-2}
  I.F. Ginzburg, G.L. Kotkin, V.G. Serbo and V.I. Telnov, Nucl. Instrum. Meth. {\bf 205} (1983) 47.

\bibitem{LHT-ph1} 
  A. Belyaev, C.-R. Chen, K. Tobe and C.-P. Yuan, Phys. Rev. {\bf D74} (2006) 115020.

\bibitem{LHT-ph21} 
  D. Choudhury, D.K. Ghosh and S.K. Rai, JHEP {\bf 07} (2012) 013.

\bibitem{LHT-ph22} 
  D. Choudhury and D.K. Ghosh, JHEP {\bf 08} (2007) 084.

\bibitem{LHT-ph23} 
  S. Mukhopadhyay, B. Mukhopadhyaya and A. Nyffeler, JHEP {\bf 05} (2010) 001.

\bibitem{LHT-ILC} 
  A.B. Mahfoudh, L. Guo, W. Liu, W.-G. Ma, R.-Y. Zhang and W.-J. Zhang, Commun. Theor. Phys. {\bf 62} (2014) 824.

\bibitem{LHT-ph41} 
  R.-Y. Zhang, H. Yan, W.-G. Ma, S.-M. Wang, L. Guo and L. Han, Phys. Rev. {\bf D85} (2012) 015017.

\bibitem{LHT-ph42} 
  X.-D. Yang, S.-J. Xiong, W.-G. Ma, R.-Y. Zhang, L. Guo and X.-Z. Li Phys. Rev. {\bf D89} (2014) 014008.

\bibitem{KLN-1}
  T. Kinoshita, J. Math. Phys. {\bf 3} (1962) 650.

\bibitem{KLN-2}
  T.D. Lee and M. Nauenberg, Phys. Rev. {\bf 133} (1964) B1549.

\bibitem{tcpss}
  B.W. Harris and J.F. Owens, Phys. Rev. {\bf D65} (2002) 094032.

\bibitem{Du} 
  S.-M. Du, L. Guo, W. Liu, W.-G. Ma and R.-Y. Zhang, Phys. Rev. {\bf D86} (2012) 054027.

\bibitem{Stefan}
 R.K. Ellis and G. Zanderighi, JHEP {\bf 02} (2008) 002.

\bibitem{OneTwoThree}
  G. 't Hooft and M. Veltman, Nucl. Phys. {\bf B153} (1979) 365.

\bibitem{Four}
  A. Denner, U. Nierste and R. Scharf, Nucl. Phys. {\bf B367} (1991) 637.

\bibitem{PDG2012}
J. Beringer, \textit{et al}. (Particle Data Group), Phys. Rev. {\bf D86} (2012) 010001.

\bibitem{LHTlimits}
  J. Reuter, M. Tonini and M. de Vries, JHEP {\bf 02} (2014) 053.

\bibitem{Cacciari:2008gp}
  M. Cacciari, G.P. Salam and G. Soyez, JHEP {\bf 04} (2008) 063.

\bibitem{jet-algorithm}
 G.P. Salam, Eur. Phys. J. {\bf C67} (2010) 637.

\bibitem{Cacciari:2011ma}
  M. Cacciari, G.P. Salam and G. Soyez, Eur. Phys. J. {\bf C72} (2012) 1896.

\end{thebibliography}
\end{document}